\documentclass[%
 reprint,
 amsmath,amssymb,
 aps,
pre,
]{revtex4-1}

\usepackage{graphicx}
\usepackage{dcolumn}
\usepackage{bm}
\usepackage{hyperref}
\usepackage[mathlines]{lineno}
\usepackage{multirow, makecell,color,gensymb}
\usepackage[english]{babel}
\usepackage{longtable}

\begin{document}

\preprint{APS/123-QED}

\title{An improved boundary condition\\ at a low grid resolution and Reynolds number}

\author{T. Burel}
 \email{thomas.burel@simuni.fr}
 \altaffiliation[Also at ]{James Weir Fluids Laboratory, Mechanical and Aerospace Department, University of Strathclyde.}
\author{Q. Gu}%
\affiliation{%
 University of Strathclyde\\
 75 Montrose Street, Glasgow.
}%


%

\date{\today}

\begin{abstract}
Complex geometries can be easily treated using the well-known full-way and half-way bounce-back rules. However, the accuracy of the full-way bounce-back rule is one order lower than the half-way bounce-back rule. Moreover, when the walls are not aligned with the lattices, the errors increase. Including the collision operator on the walls with the full-way bounce-back rule leads to an improvement of the accuracy of the pressure-drop, but, a loss of momentum is observed on concave corners. We propose to improve the momentum conservation by adding an extrapolation of the density by the inverse distance weighting at concave corners. The technique is shown to give a second-order accuracy at a lower grid resolution, thus, the computational cost can be reduced.
\end{abstract}

\pacs{Valid PACS appear here}
\maketitle

\section{Introduction}
\paragraph*{}Optimisation of existing system and designing new generation of products with compactness in mind, it is a challenge. Achieving this kind of compactness is mostly carried out by utilising porous media such as membranes, foams, etc. Therefore, the needs of simulation for complex geometries have increased drastically. At the same time, the accuracy of simulation has increased, thus, the simulations became more predictive. Different methods have been developed for Computational Fluid Dynamics such as Lattice Boltzmann Method (LBM) which is a mesoscopic particle-based method derived from Boltzmann equation and Lattice Gas Automata (LGA) \cite{Hardy1973}. The collision between particles can be approximated by different operators such as Bhatnagar-Gross-Krook (BGK) \cite{Bhatnagar1954}, the Multiple Relaxation Time (MRT) \cite{DHumieres2002,Lallemand2000}, the Central Moment (CM) \cite{Geier2009,Premnath2009}, the entropic \cite{Dorschner2017,Bosch2015}, etc. 
\paragraph*{} A large branch of wall boundary conditions has been developed. The full-way bounce-back rule is the simplest rule, in which, each particle that hits the wall bounces in the opposite direction instantaneously without being affected by a collision process. This rule ensures the conservation of mass and the no-slip condition at the boundary. Removing the collision operator at the wall nodes reduces drastically the accuracy which results in great demand of finer mesh. An extension has been done by considering the wall is located between two lattice nodes. Using the approximation of Chapman-Enskog expansion, the full-way and half-way are first and second order, respectively \cite{Ginzbourg1996}.\\
The half-way bounce back rule can lead to a slip velocity at the wall. To tackle this problem, several wall boundary conditions were proposed such as Inamuro \textit{et al} in 1995 \cite{Inamuro1995}  which constructs a reflective and diffuse boundary by using the local equilibrium function, Chen \textit{et al} in 2007 \cite{Chun2007} proposed to interpolate the equilibrium distribution, Latt in 2007 \cite{Latt2008,LattThesis2007} built a scheme based on the regularised LBM, Ginzburg \cite{Ginzbourg1996}, Ladd \cite{Ladd1994}, and Bouzidi \cite{Bouzidi2001} proposed to interpolate schemes. 

\paragraph*{} Choosing LBM to treat porous media can cause a misalignment  between the geometry (walls) and the Cartesian grid (lattice) \cite{Kruger2003}. Enforcing the alignment leads to the generation of ``stairs'' with introduces concave and convex corners. Using a classical scheme for the Lattice Boltzmann Method, the density at a concave corner node is not correctly calculated since the distribution is not completely defined.
Thus, it leads to having sensible different pressure with the neighbour nodes at the same time step. We proposed to extrapolate the density in a suitable manner to improve the accuracy and the conservation of the momentum for flows at a low Reynolds number.

\section{Methodology}
\subsection{LBM scheme} 
Considering a Cartesian grid and assuming the particles cannot be off-lattice, the LBM equation can be written as 
\begin{equation}
f_i(\vec{x}+\vec{e}_i \Delta x,t+\Delta t)=f_i(\vec{x},t)+\Omega_i(f_i(\vec{x},t))  + F_i(\vec{x},t) ,
\end{equation}
where $\Omega_i(f_i(\vec{x},t)) $ is the LBM collision operator and $F_i(\vec{x},t)$ is an external force.
The collision operator is modelled by the well-known BGK in the linear form to simplify and reduce the cost of the collision term. Applying this operator to LBM, we get the LBGK model \cite{QIAN1990,Chen1991,Qian1992} and is written as
\begin{equation}\label{LBGK}
\begin{split}
f_i(\vec{x}+\vec{e}_i \Delta x,t+\Delta t) = f_i(\vec{x},t) &- \frac{f_i(\vec{x},t)-f_i^{eq}(\vec{x},t)}{\tau}  \\ &+ F_i(\vec{x},t) ,
\end{split}
\end{equation}\\
where $\tau$ is the relaxation factor and defined as
\begin{equation}\label{relaxation-factor}
\tau = \frac{6 \cdot \nu + 1}{2} ,
\end{equation}\\
where $\nu$ is the kinematic viscosity and $f_i^{eq}$ is the equilibrium distribution.
Considering D2Q9 scheme (two-dimensional square lattice with 9 discrete velocities) \cite{Qian1992}, the Maxwell-Boltzmann equilibrium distribution is:
\begin{equation}\label{feq-D2Q9}
f_i^{eq}=\rho\omega_i \bigg[ 1+3 \vec{e}_i\cdot \vec{u}+\frac{9}{2} \big( \vec{e}_i \cdot \vec{u} \big)^2-\frac{3}{2}\vec{u}^2 \bigg] ,
\end{equation}.\\
where, $e_0=c (0,0)$, $e_1=c (1,0)$, $e_2=c (0,1)$, $e_3=c (-1,0)$, $e_4=c (0,-1)$, $e_5=c (1,1)$, $e_6=c (-1,1)$, $e_7=c (-1,-1)$, $e_8=c (1,-1)$ and $w_0=4/9$, $w_1=w_2=w_3=w_4=1/9$, $w_5=w_6=w_7=w_8=1/36$. The density $\rho$ is calculated by the zeroth-moment ($\sum f_i$) and the macroscopic velocity $\vec{u}$ is the first-moment ($\sum f_i \cdot \vec{e}_i$). $c$ is the lattice speed (set to 1) and the speed of sound $c_s$ is $1/\sqrt{3}$ in the simulations. Thus, the pressure is $P=\rho * c_s^2$ by using the ideal gas law.

\subsection{Boundary treatment} 
\begin{figure}[htbp]
  \begin{center}
   \includegraphics[width=0.4\textwidth]{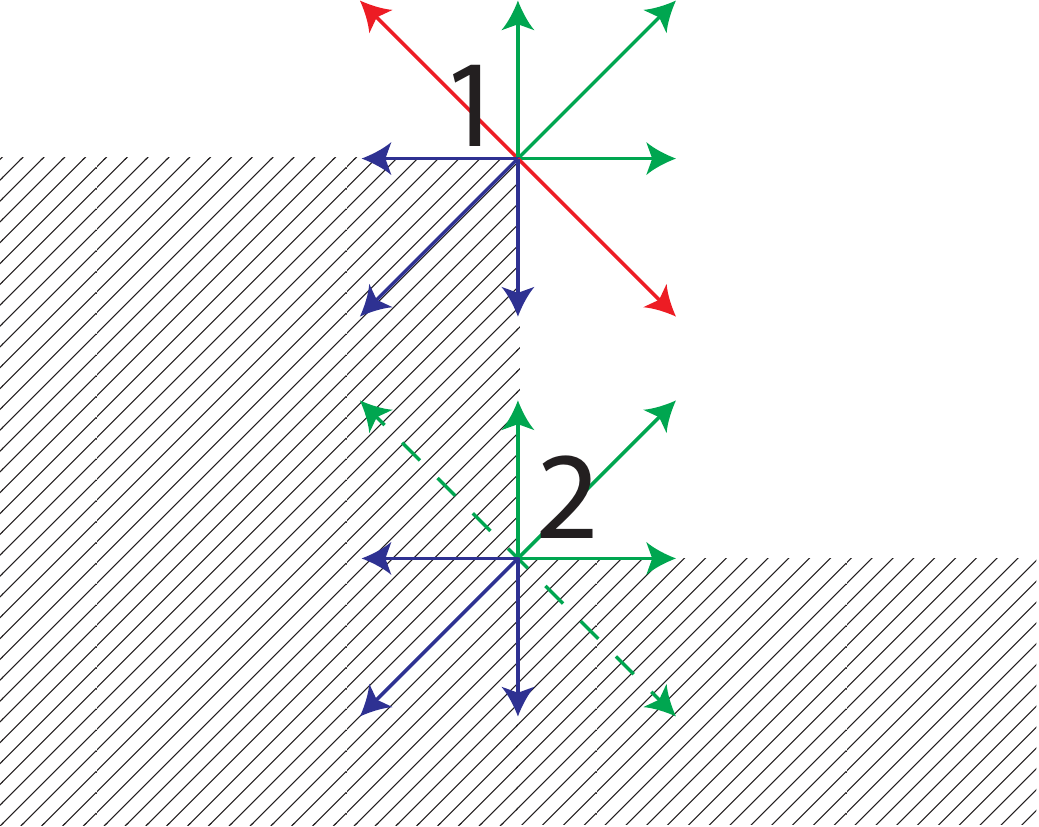} 
\caption{Simplify representation of the D2Q9 distribution at convex (1) and concave (2) corners. Blue and green lines are the incoming and outgoing discrete particles, respectively. The red lines represent the incoming and the outgoing discrete particles. The green dash lines are the unknown discrete particles.}
  \label{fig:schema_corners}
  \end{center}
\end{figure}

All the walls aligned with the lattice are treated with the full-way bounce-back rules with streaming in all direction excluding towards the solid nodes and the full BGK collision including towards the solid nodes is applied after the bounce-back rule. The corners are treated in a relatively similar manner to wall boundary conditions, however, at concave corners, the two discrete particle velocities pointing into the solid in opposite directions represented by the dash lines in \autoref{fig:schema_corners} are unknown. These two discrete particle velocities do not participate in the streaming process but affect the collision process in the D2Q9 schemes i.e. the system of equations at the concave corners are ill-posed. Thus, some choice has to be made to close the system. The simplest approach is to set these to zero but this leads to a loss of momentum. Another approach is to extrapolate the distribution function at concave corners from the neighbouring nodes \cite{Kruger2003}. In the present work, the set of equation is closed, specifically for low Reynolds flows, by a macroscopic approach. The assumption of low Reynolds number implies that  the diffusive time scale is much greater than the convective ones i.e. the inertia forces are negligible compared to the viscous forces. 
This enables us to consider the gradient of density is small. Thus, the density at the concave corner is extrapolated by the inverse distance weighting of the three direct neighbours and the two unknown discrete particles are defined by using the zeroth-moment and assuming the two unknown discrete particles equal, as
\begin{equation}
2 f_{un}=\tilde{\rho}- (2 \rho_{in} + f_0) ,
\end{equation}
where, $f_{un}$, $\tilde{\rho}$, $\rho_{in}$, $f_0$ are the unknown discrete particle, the extrapolate density, the sum of incoming discrete particles, and the static discrete particle, respectively.%
In this way, the density at the concave corner is approximated to a realistic value which produces quicker and better simulations.\\
To define the extrapolation, let's consider the density at the concave corner $\tilde{\rho}$ and n nodes. The subscripts $i$ and $j$ refer to the indexes in the directions x and y, respectively in this section.
The general form of the extrapolation of the density is:
\begin{equation}
\tilde{\rho}\approx\frac{\sum\limits^{n}_{k=1}\omega_{k}\rho_{k}}{\sum\limits^{n}_{k=1}\omega_{k}},
\end{equation}
where,
\begin{equation}
\omega_{k}=\frac{1}{\sqrt{\left(x\left(\rho_{k}\right)-x\left(\tilde{\rho}\right)\right)^2 +\left(y\left(\rho_{k}\right)-y\left(\tilde{\rho}\right)\right)^2}} .
\end{equation}
Thus, for a concave corner oriented in the +x-direction and the +y-direction, the extrapolation on the boundary is:
\begin{equation}
\tilde{\rho}\approx \frac{1}{2+\frac{1}{\sqrt{2}}} \left(\frac{1}{\sqrt{2}}\rho_{i+1,j+1}+\left(\rho_{i+1,j}+\rho_{i,j+1}\right)\right) .
\end{equation}
Due to the extrapolation uses the density values of the nearest neighbours i.e. the first layer of the surrounding nodes, the parallel efficiency is not strongly impacted. 
The two incoming and outgoing discrete particles represented by the red lines in \autoref{fig:schema_corners} are assumed equal to enforce the no-slip boundary condition, thus, the incoming discrete particles are summed and are equal to the sum of the outgoing discrete particles to conserve the mass and momentum.
\section{Simulation conditions}
\begin{figure}[htbp]
  \begin{center}
   \includegraphics[width=0.4\textwidth]{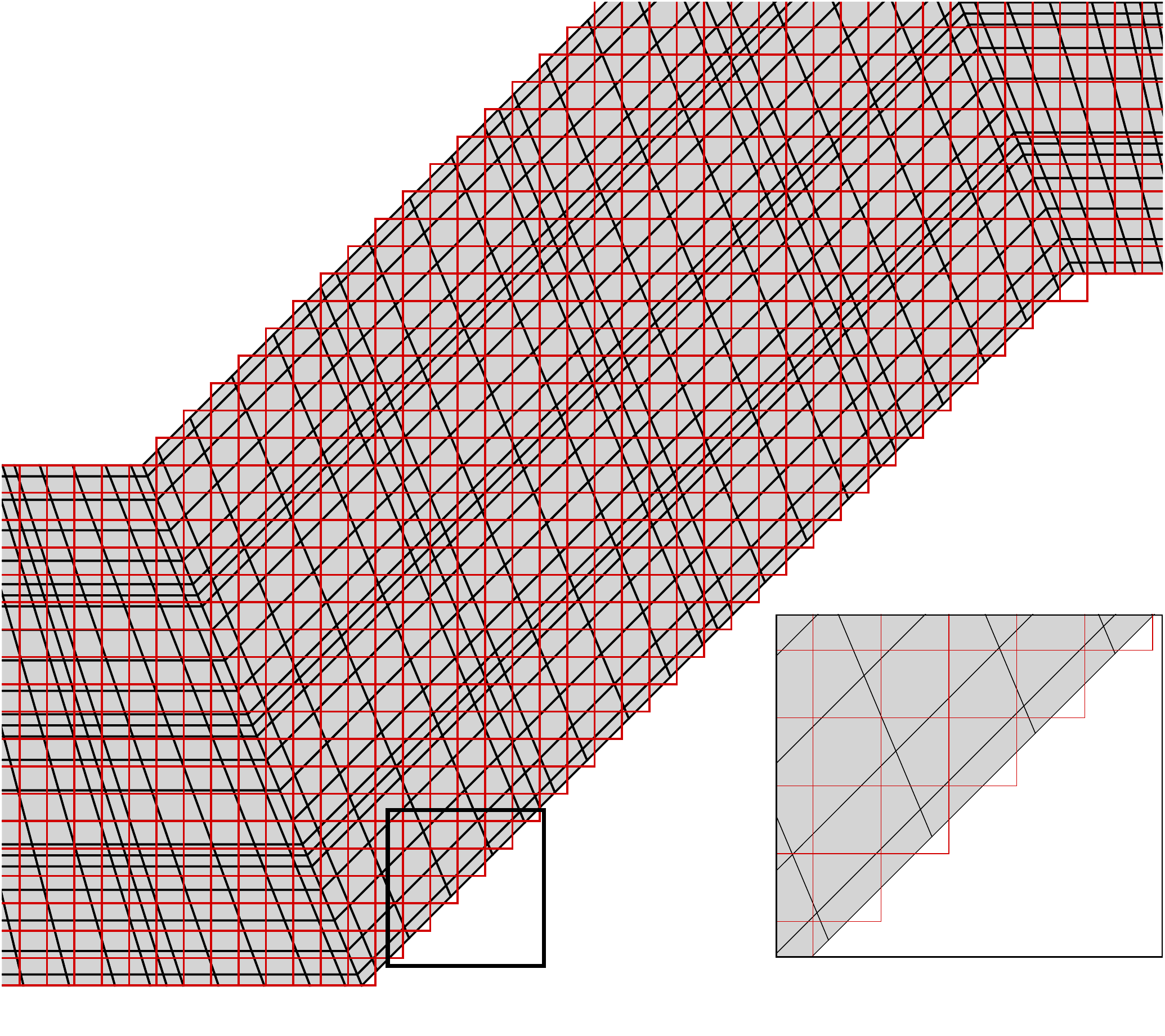} 
\caption{Comparison of the LBM (Cartesian) and SEM meshes in red and black, respectively.}
  \label{fig:Mesh-match}
  \end{center}
\end{figure}

\begin{figure}[htbp]
  \begin{center}
   \includegraphics[width=0.4\textwidth]{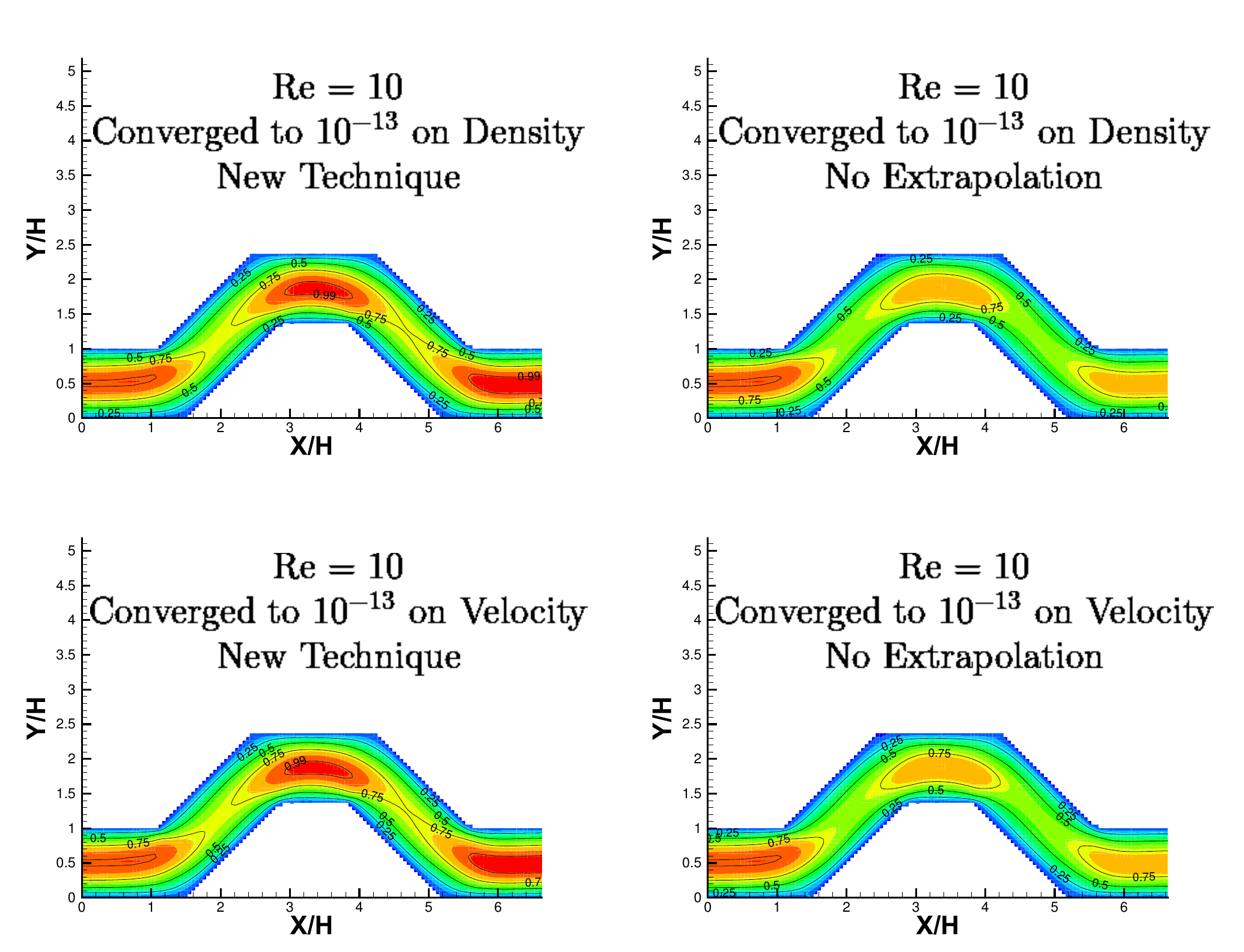} 
\caption{Comparison of the x-velocity field for Re=10 and low grid resolution}
  \label{fig:Inclined-Valid-Velocity}
  \end{center}
\end{figure}

We define the boundary conditions treated with the standard full-way bounce-back as ``No Collision ''. The new approach with the collision operator and the extrapolation of the density from the neighbour nodes by the inverse distance weighting approach is called the ``New Technique'', whereas, without extrapolation the ``No Extrapolation''. Therefore, for walls aligned with the lattice, the ``New Technique'' and ``No Extrapolation'' are the same and are called ``With Collision''. We also compared to the standard half-way bounce-back and named ``Half-Way''. To analyse the new treatment, we use a Poiseuille-like flow in straight and inclined channels. The inclined cases are compared to Spectral Element Method (SEM) with Nek5000 \cite{Fischer2005}. Four elements with the polynomial order of 7 in the cross-section of the channels, the $P_n/P_{n-2}$ formulation, the Helmholtz solver convergence criterion of $10^{-12}$, and the divergence criterion of $10^{-11}$ were used for the SEM simulations. The mesh for the SEM simulations cannot be the same as LBM. In order to have the simulations as close as possible, we keep the same volume of fluid and the straight walls are aligned, thus the walls for the stairs are located in the middle as shown in \autoref{fig:Mesh-match}. The viscosity in LBM simulations is set to 0.1 [lu]. The pressure-drop is calculated by the difference of pressure between the centre of the channels at the inlet and the outlet imposed pressure. We run the simulations for inclined channels with two grid resolutions and compared to the straight channels. In case of high grid resolution, the height of the channel is 192 lattices and in case of low resolution, the height of the channel is 19 lattices.
\paragraph*{}
The Reynolds number is defined as
\begin{equation}\label{Eq:Reynolds-Poiseuille-Single}
Re=\frac{U^{max} H}{\nu} ,
\end{equation}
where, $U^{max}$ is the maximum velocity in the channel, $H$ the height of the channel and $\nu$ the kinematic viscosity.
\paragraph*{}
In \autoref{fig:Inclined-Valid-Velocity}, the velocity field is represented for two conditions of convergence: one based on the density and one on the X component of the velocity. Both criteria lead to essentially the same results.
In the following, the calculations will be terminated when the evolution of density is lower than $10^{-13}$  or $10^{-15}$ over the domain for low and high grid resolution, respectively. Indeed, the convergence criterion needs to be smaller for high grid resolution due to the velocity magnitude becomes extremely small, thus, the pressure-drop too which is the important result of this study. Those criteria were chosen after a sensibility study.
\section{Straight channels}
\subsection{Validation}
\begin{figure}[h!]
  \begin{center}
  	\scalebox{0.4}{%
  	\includegraphics[width=1.0\textwidth,trim={5.0cm 14.5cm 4.5cm 4.5cm},clip]
  	{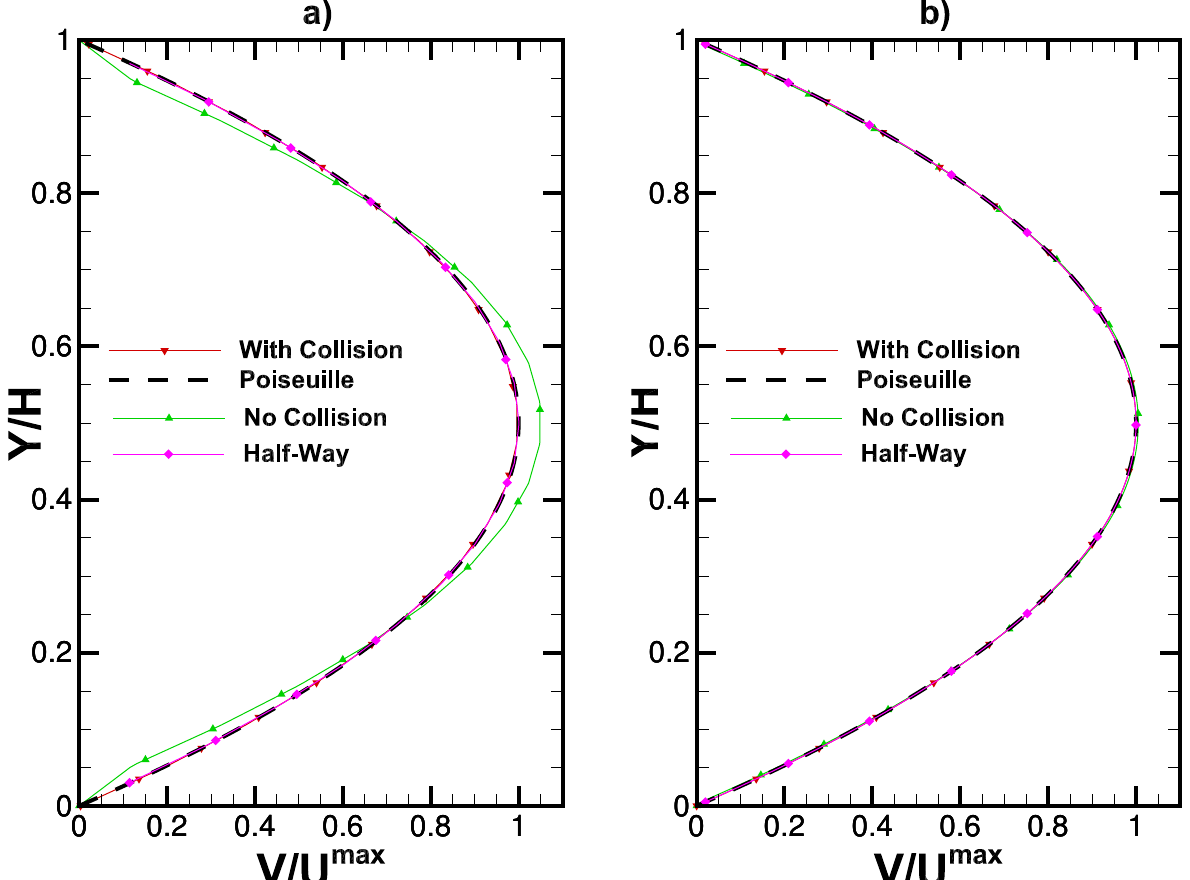}
} 
\caption{Velocity profiles at the outlet: a) low grid resolution and Re = 0.01 and b) high grid resolution and Re = 1.}
  \label{fig:Straight-Vel-profiles-Outlet}
  \end{center}
\end{figure}

The straight channels validate our code for a well-known flow with an analytical solution which is
\begin{equation}\label{Eq:Poiseuille-Single}
\Delta P=\frac{8Re\nu^{2}}{H^{3}}L ,
\end{equation}
where, $H$ and $L$ are the height and  length of the channel, respectively. This length is equal to two times the height of the channel.\\ 
A parabolic velocity profile is imposed at the inlet on the left side of the channel and a constant pressure set to 1/3 [lu] on the right side of the channel with Zou and He  boundary condition \cite{Zou1997}.
\paragraph*{Pressure-drop analysis}
\autoref{tab:Poiseuille-Straight} shows the collision on the wall is needed to get a more accurate pressure-drop. The error is calculated with the analytical solution as a reference and normalised with the analytical solution. 
\begin{table}[htbp]
\centering
\begin{ruledtabular}
\begin{tabular}{ccccc}
\multicolumn{5}{c}{\textbf{Straight channels}}                                                                                        \\ \hline
\multicolumn{1}{l}{}  & \textbf{Re} & \thead{\textbf{With}\\ \textbf{Collision}} & \thead{\textbf{No}\\ \textbf{Collision}} & \thead{\textbf{Half-}\\ \textbf{Way}} \\ \hline
\multirow{4}{*}{\thead{\textbf{Low}\\ \textbf{Resolution}}}  
& 10    & 1.47\%    & 18.75\%     & 2.19\%    \\ 
& 1     & 0.28\%    & 14.43\%     & 0.33\%    \\ 
& 0.1   & 0.39\%    & 14.06\%     & 0.20\%    \\ 
& 0.01  & 0.41\%    & 14.02\%     & 0.19\%    \\ \hline
\multirow{4}{*}{\thead{\textbf{High}\\ \textbf{Resolution}}} 
& 10    & 0.014\%    & 1.507\%     & 0.021\%  \\ 
& 1     & 0.003\%    & 1.285\%     & 0.003\%  \\ 
 & 0.1  & 0.004\%    & 1.249\%     & 0.002\%  \\ 
& 0.01  & 0.007\%    & 1.247\%     & 0.004\%  \\ 
\end{tabular}
\end{ruledtabular}
\caption{Pressure-drop errors for a Poiseuille flow in the straight channels}
\label{tab:Poiseuille-Straight}
\end{table}
In \autoref{tab:Poiseuille-Straight}, it can be noticed the grid size is not enough for the low-resolution geometry to have an error less than 1\% for Re=10. Compared to the classical ``No Collision '' technique, the results of using the collision operator on the walls (the ``With Collision'') are in good agreement with the analytical solution and similar to the ``Half-Way'' results.
\paragraph*{Velocity profile analysis}
We have extracted the velocity profiles at the outlet which are shown in \autoref{fig:Straight-Vel-profiles-Outlet}. The Reynolds number has not strongly modified the velocity profiles for the creeping flow regime. It can also be noticed that the incorrect wall shear rate for the ``No Collision'' and thus, an overshoot of momentum is observed. 

\subsection{Convergence study}
The simulations were carried out for the channel height from 4 to 256 lattices, for ``No Collision'', ``With Collision'', and ``Half-Way'' wall boundary conditions, and for Reynolds numbers of 0.01, 0.1, 1, and 10. However, the results for Re= 0.1 or 1 are similar to 0.01, thus, the results for Re=0.1 and 1 are not shown.
\subsubsection{Study of the errors versus channel height}
\begin{figure}[h!]
  \begin{center}
    \includegraphics[width=0.4\textwidth,trim={0.3cm 0.3cm 0.3cm 0.3cm},clip]
    {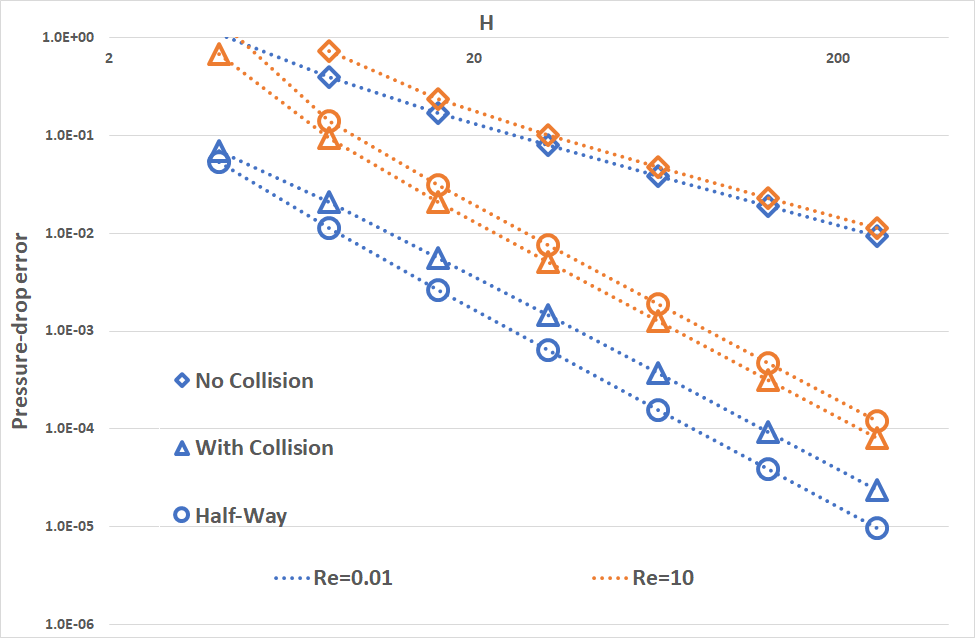} 
\caption{The pressure-drop error versus channel height.}
  \label{fig:Straight-pressure-drop}
  \end{center}
\end{figure}

\begin{figure}[h!]
  \begin{center}
    \includegraphics[width=0.4\textwidth,trim={0.3cm 0.3cm 0.3cm 0.3cm},clip]
    {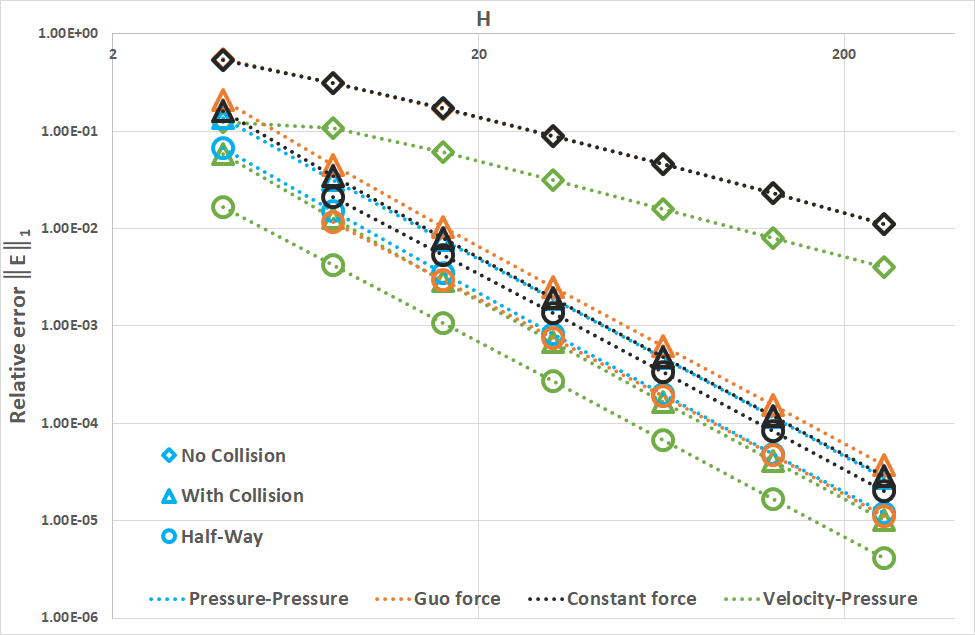} 
\caption{The relative error of velocity at Re=0.01 versus channel height.}
  \label{fig:Straight-Velocity-Error-Re0-01}
  \end{center}
\end{figure}

\begin{figure}[h!]
  \begin{center}
    \includegraphics[width=0.4\textwidth,trim={0.3cm 0.3cm 0.3cm 0.3cm},clip]
    {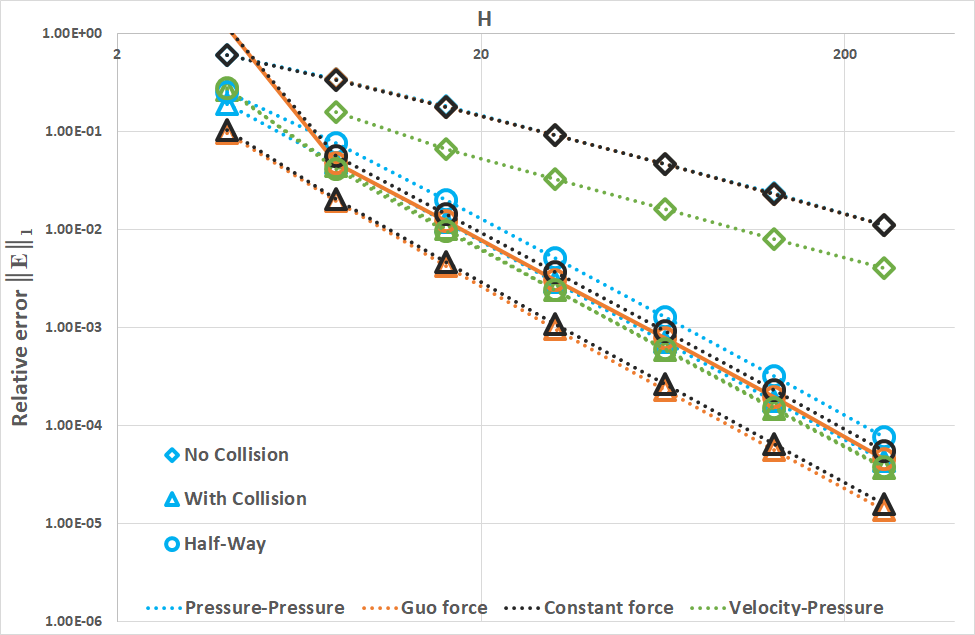} 
\caption{The relative error of velocity at Re=10 versus channel height.}
  \label{fig:Straight-Velocity-Error-Re10}
  \end{center}
\end{figure}

\paragraph{Pressure-drop errors}
The \autoref{fig:Straight-pressure-drop} shows ``No Collision'' converges to the same error for Re=0.01 or 10 whereas ``With Collision'' or ``Half-Way'' gives more accurate results for lower Reynolds number. Moreover, ``Half-Way'' gives more accurate results than ``With Collision'' when the Reynolds is lower.
\paragraph{Velocity errors}
We have performed the calculus of the relative error $\left\Vert E\right\Vert_1$ of the velocity  as
\begin{equation}\label{eq:relative-error}
\left\Vert E\right\Vert_1 =\frac{\sum^N|U_{c}-U_{a}|}{\sum^N |U_{a}|} ,
\end{equation}
where, N is the number of points in the domain, $U_c$ and $U_a$ are the velocity along the channel calculated from the LBM results and the analytical solution, respectively.\\
The Poiseuille-like flow can be imposed in a number of ways. In the current study, it is imposed by a constant force in the domain \cite{He1997} (CF) i.e. $F_i = 3  \omega_i e_{i} \cdot (\partial P/\partial x)$, the Guo source term \cite{Guo2002} (GF) i.e. $F_i = \Delta t \ \omega_i \left(1-1/(2\tau) \right) \left[((u \cdot e_i)/c_s^4)e_i +(e_i-u)/ c_s^2\right]\cdot (\partial P/\partial x)$, Pressure at inlet and outlet by Zou and He \cite{Zou1997} (PP), and as above Velocity at inlet and Pressure at outlet by Zou and He \cite{Zou1997} (VP).\\
\autoref{fig:Straight-Velocity-Error-Re0-01} shows ``No Collision'' has a lower order of convergence. PP, GF, or CF give the same results for ``No Collision'' whereas, VP gives the best accuracy. The  ``Half-Way'' technique with VP  produces the most accurate one for creeping flows (\autoref{fig:Straight-Velocity-Error-Re0-01}) but for a weak laminar flow (\autoref{fig:Straight-Velocity-Error-Re10}), it is ``With Collision'' with GF. Thus, the relative errors of velocity confirm the trend of the pressure-drop error that ``Half-Way'' is better for $Re \leqslant 1$ otherwise ``With Collision'' produce more accurate results. The \autoref{fig:Straight-Velocity-Error-Re0-01} shows GF is preferable to CF for ``Half-Way'' but the opposite for ``With Collision''. This rises from the fact GF is a body force (volumetric) whereas CF is a discrete external force. Globally, PP gives errors between GF and CF.
When the inertia force becomes non-negligible, the errors rise and especially for a channel height of 4 lattices (\autoref{fig:Straight-Velocity-Error-Re10}) due to non-linearity of the flow appears. GF and PP give the most and worst accurate, respectively.
\subsubsection{The rate of convergence study}
Figures \ref{fig:Straight-pressure-drop}, \ref{fig:Straight-Velocity-Error-Re0-01}, and \ref{fig:Straight-Velocity-Error-Re10} show a linear rate of convergence (q). Thus, we have extracted the rate of convergence between the grid resolution of 8 and 256 lattices of channel height as
\begin{equation}
q=\log_{32} \left\Vert E\right\Vert_1 (H=8)/\left\Vert E\right\Vert_1(H=256) .
\end{equation}
The results are summarized in \autoref{tab:Rate-Convergence}. It is observed that the ``No Collision'' converges as a first-order scheme whereas, ``With Collision'' or ``Half-Way'' converges as a second order scheme. As expected, the collision is needed to obtain higher order rate of convergence and achieves a second-order accuracy. 
\begin{table}[htbp]
  \centering
  \begin{ruledtabular}
    \begin{tabular}{cccccc}
    \multicolumn{6}{c}{\textbf{Rate of convergence}} \\
\hline    Re &     & \thead{\textbf{Pressure-} \\ \textbf{Pressure}} & \thead{\textbf{Guo} \\ \textbf{force}} & \thead{\textbf{Constant} \\ \textbf{force}} & \thead{\textbf{Velocity-} \\ \textbf{Pressure}} \\
    \hline
    \multirow{3}{*}{0.01} & No Collision & 0.96  & 0.96  & 0.96  & 0.95 \\
     & With Collision & 2.03  & 2.05  & 2.05  & 2.06 \\
    & Half-Way & 2.06  & 2.00  & 2.01  & 2.00 \\
    \hline
    \multirow{3}{*}{10} & No Collision & 0.99  & 0.99  & 0.98  & 1.06 \\
    & With Collision & 2.03  & 2.09  & 2.07  & 2.05 \\
    & Half-Way & 1.99  & 2.00  & 2.00  & 2.02 \\

    \end{tabular}%
   \end{ruledtabular}
    \caption{The rate of convergence for straight channels.}
  \label{tab:Rate-Convergence}%
\end{table}%
\section{Inclined channels}\label{sec:inclined-channels}
\begin{figure}[htbp]
  \begin{center}
    \includegraphics[width=0.45\textwidth,trim={0.3cm 0cm 0.3cm 0cm},clip]
    {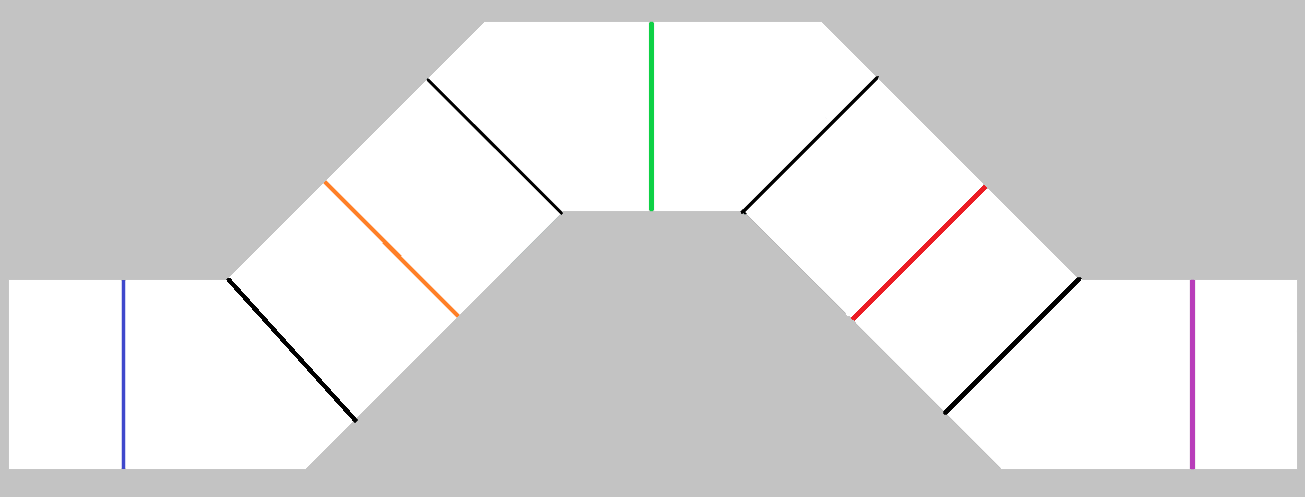}  
\caption{The geometry of inclined channels where the dark blue, orange, green, red, and purple lines represent the stations 1,2,3,4, and 5, respectively. The black lines represent the limit of the inclined channels.}
  \label{fig:Inclined-geom}
  \end{center}
\end{figure}

\paragraph*{}
This case was designed in order to have a channel with two sections inclined by 45\degree and 5 equal ``X'' lengths of the centre line of the channel. Design Modeler from ANSYS \cite{AnsysInc.2016} was used to design it and an image was exported then Matlab from MathWorks \cite{MathWorks2014} was used to scale the image and export to a binary format.\\
The original image before converting to a binary file at the right scale can be seen in \autoref{fig:Inclined-geom}. However, this geometry cannot keep the same mesh and the same height of channel for full-way and half-way bounce-back treatments. We chose to keep the same mesh to compare our results, thus, the heights of channel are 193 and 20 lattices for the ``Half-Way''. 
\subsection{Effect on velocity and momentum}
The accuracy of the velocity profile is critical in case of additional transport properties such as chemical species or light solid particles. In two phases flow, the mass fraction of one fluid needs to be transported. Thus, the inaccuracy of the velocity profile will affect the position of the interface when the capillary number becomes high. Moreover, dealing with a complex geometry and a big domain do not give the opportunity to use a fine grid in each channel. A previous study mentioned at least 4 lattices in each channel is needed to support Poiseuille-like behaviour \cite{Succi2001}. However, we have noticed for this case it needs at least 10 lattices to have an acceptable pressure-drop error ( more or less 10\%) i.e. similar to the straight channel. 
\subsubsection{Analysis at the outlet of the inclined channel}
\begin{figure}[h!]
  \begin{center}
  	\scalebox{0.4}{%
  	\includegraphics[width=1.0\textwidth,trim={5.0cm 14.5cm 4.5cm 4.5cm},clip]
  	{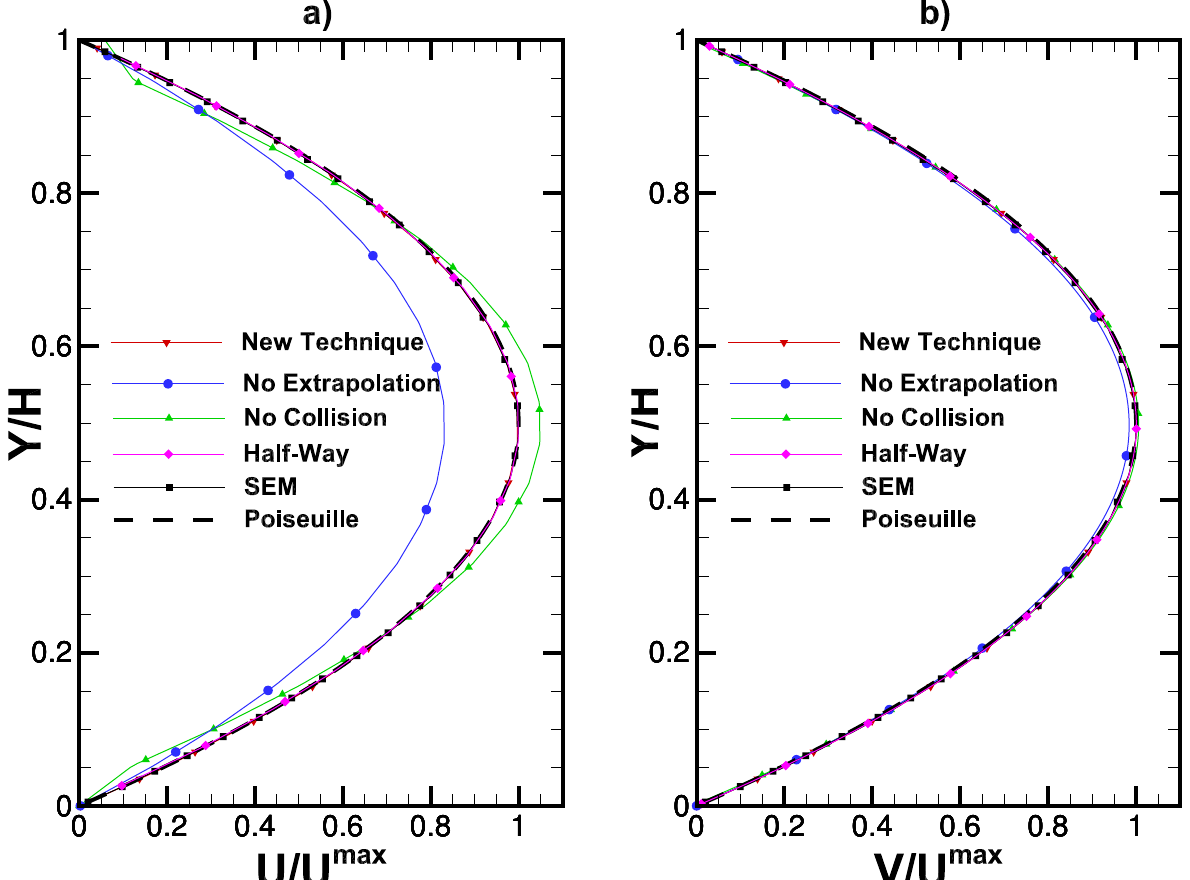}  	
}    
\caption{Velocity profiles at the outlet: a) low grid resolution and Re = 0.01 and b) high grid resolution and Re = 1.}
  \label{fig:Inclined-Vel-profiles}
  \end{center}
\end{figure}

\begin{figure}[h]
  \begin{center}
    \includegraphics[width=0.4\textwidth]
    {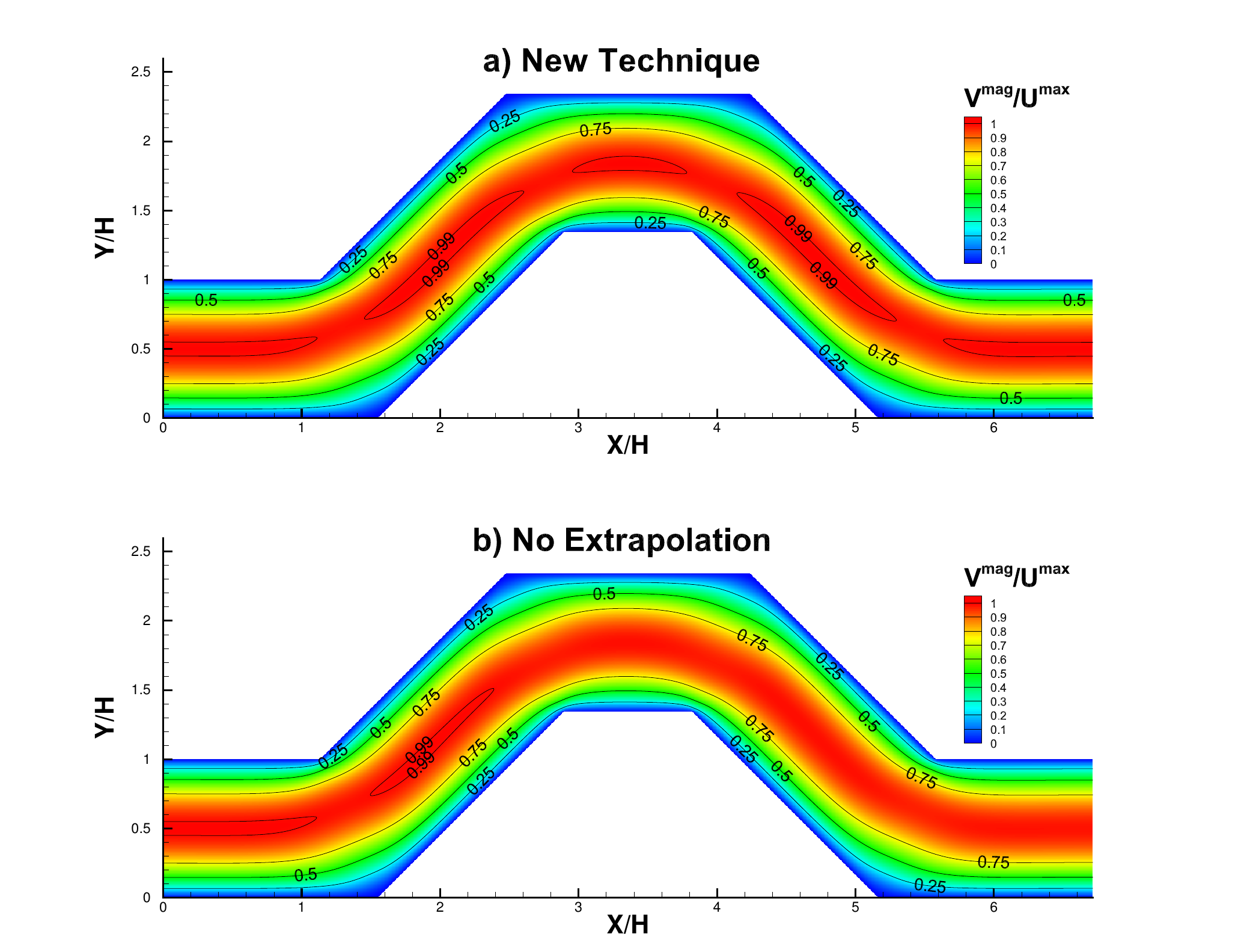}  
\caption{Velocity field for high grid resolution}
  \label{fig:Inclined-Vel-field-HighReso}
  \end{center}
\end{figure}

\begin{figure}[h]
  \begin{center}
    \includegraphics[width=0.4\textwidth]
    {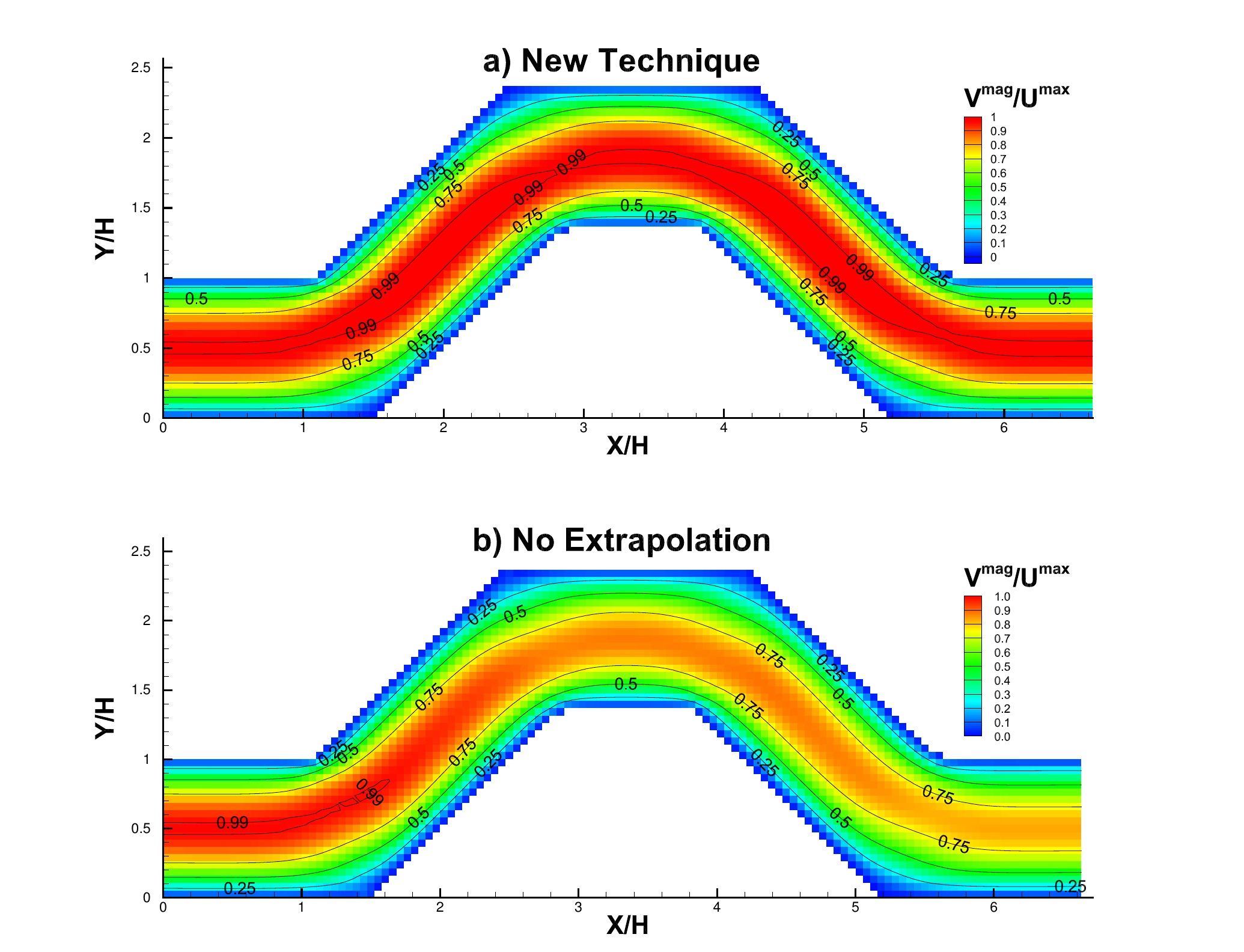}  
\caption{Velocity filed for low grid resolution}
  \label{fig:Inclined-Vel-field-LowReso}
  \end{center}
\end{figure}

In this first analysis, we seek to obtain the parabolic velocity profile at the outlet for Reynolds number less than 1.\\
In \autoref{fig:Inclined-Vel-profiles}b, for high grid resolution, the outlet velocity profiles show the ``New Techniques'', ``No Collision'', and ``Half-Way'' are able to get the velocity profile in excellent agreement with SEM whereas, the ``No Extrapolation'' cannot provide accurate profiles. In case of low grid resolution, the ``No Collision'' becomes inaccurate as seen in \autoref{fig:Inclined-Vel-profiles}a.
\\
The momentum is correctly conserved for the ``New Technique'', ``No Collision'', and ``Half-Way'' approaches. However, a loss of momentum conservation appears with the ``No Extrapolation''. Using the high grid resolution, the loss of momentum is relatively small as it can be seen in Figures \ref{fig:Inclined-Vel-profiles}b and \ref{fig:Inclined-Vel-field-HighReso}. 
In case of low grid resolution, the serious loss of momentum conservation is observed with the ``No Extrapolation'' technique as seen in Figures \ref{fig:Inclined-Vel-profiles}a and \ref{fig:Inclined-Vel-field-LowReso}. By a trapezoidal integration, the loss of momentum for the ``No Extrapolation'' is more than 10\% but for the other techniques, it is less than $10^{-3}\%$. Therefore, the ``New Technique'' improves the velocity profile in case of low grid resolution. 
Moreover, the normalised velocity profile is not affected by the Reynolds number as expected for creeping flows.

\subsubsection{Analysis in the interior of the inclined channel}
We have seen the ``New Technique'' conserves the momentum and the velocity profile is correctly captured at the outlet. However, we need an accurate velocity profile everywhere in the domain, thus, we have extracted the velocity profiles in 5 stations as shown in \autoref{fig:Inclined-geom} and have calculated $\left\Vert E\right\Vert_1$ i.e. the relative error norm as in \autoref{eq:relative-error} where N is the number of extracted points and $U_c$ and $U_a$ are the magnitude velocity calculated from the LBM results and the analytical solution, respectively.\\

\begin{figure}[h!]
  \begin{center}
  	\scalebox{0.4}{%
  	  	\includegraphics[width=1.0\textwidth,trim={5.0cm 14.5cm 4.5cm 4.5cm},clip]
  	{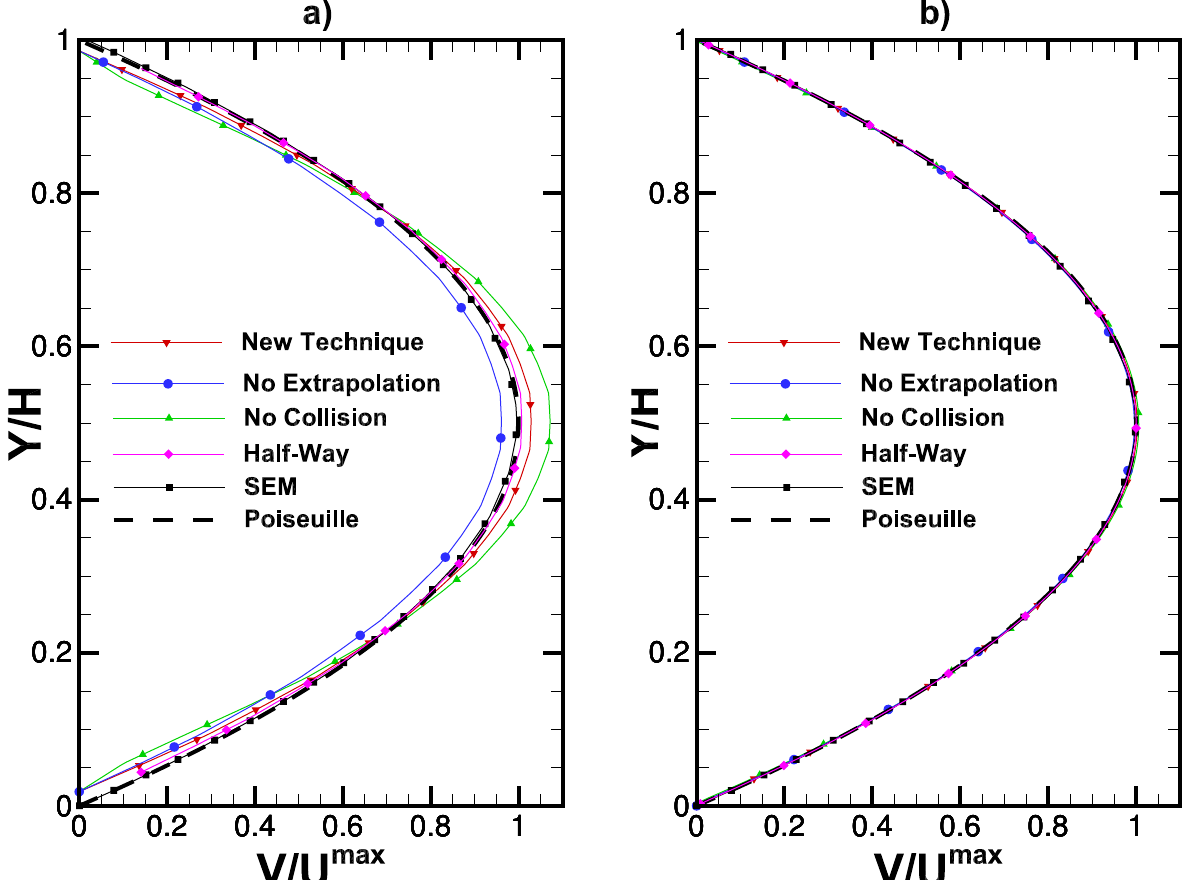} 
}    
\caption{Velocity profiles at station 2: a) low grid resolution and Re = 0.01 and b) high grid resolution and Re = 1.}
  \label{fig:Inclined-Vel-profiles-Sland1}
  \end{center}
\end{figure}

\begin{figure}[h!]
  \begin{center}
  	\scalebox{0.4}{%
  	  	  	\includegraphics[width=1.0\textwidth,trim={5.0cm 14.5cm 4.5cm 4.5cm},clip]
  	{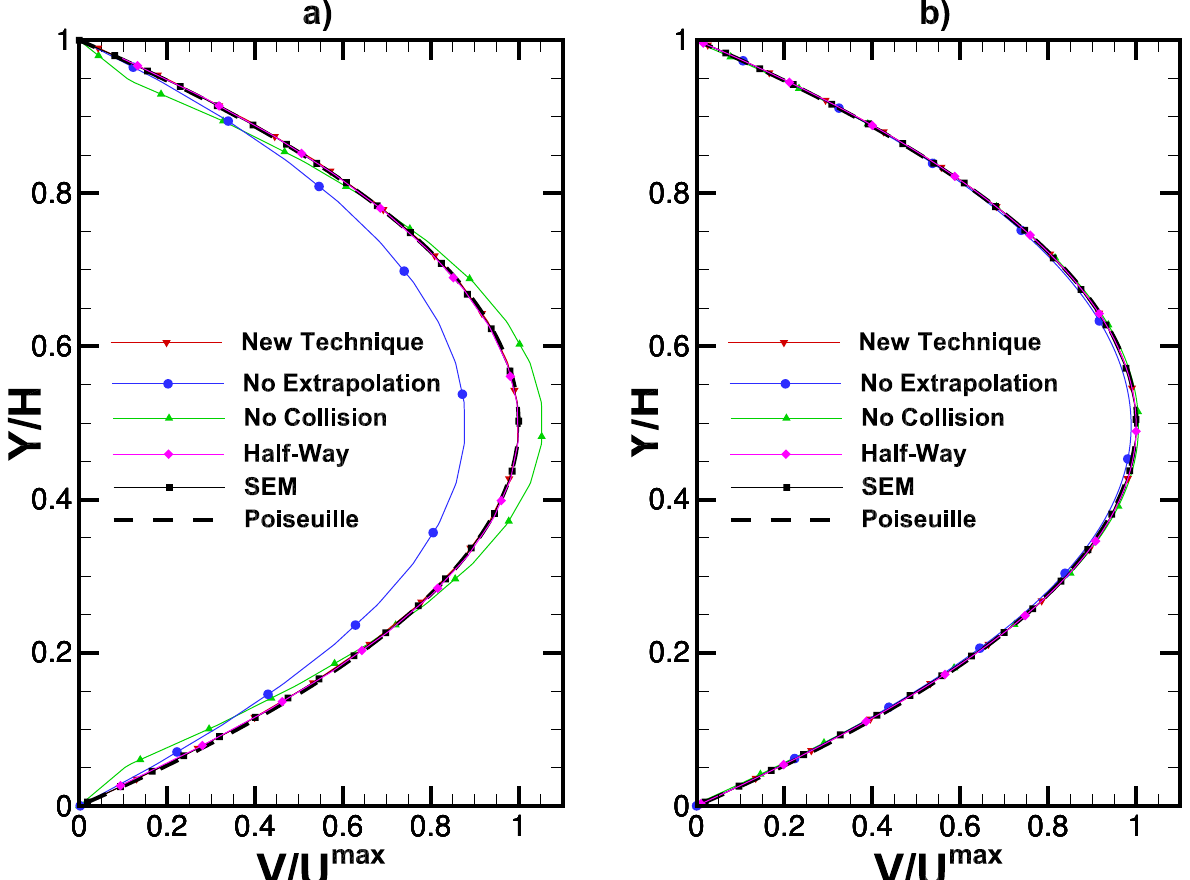} 
} 
\caption{Velocity profiles at station 3: a) low grid resolution and Re = 0.01 and b) high grid resolution and Re = 1.}
  \label{fig:Inclined-Vel-profiles-Middle}
  \end{center}
\end{figure}

\begin{figure}[h!]
  \begin{center}
  	\scalebox{0.4}{%
\includegraphics[width=1.0\textwidth,trim={5.0cm 14.5cm 4.5cm 4.5cm},clip]
    {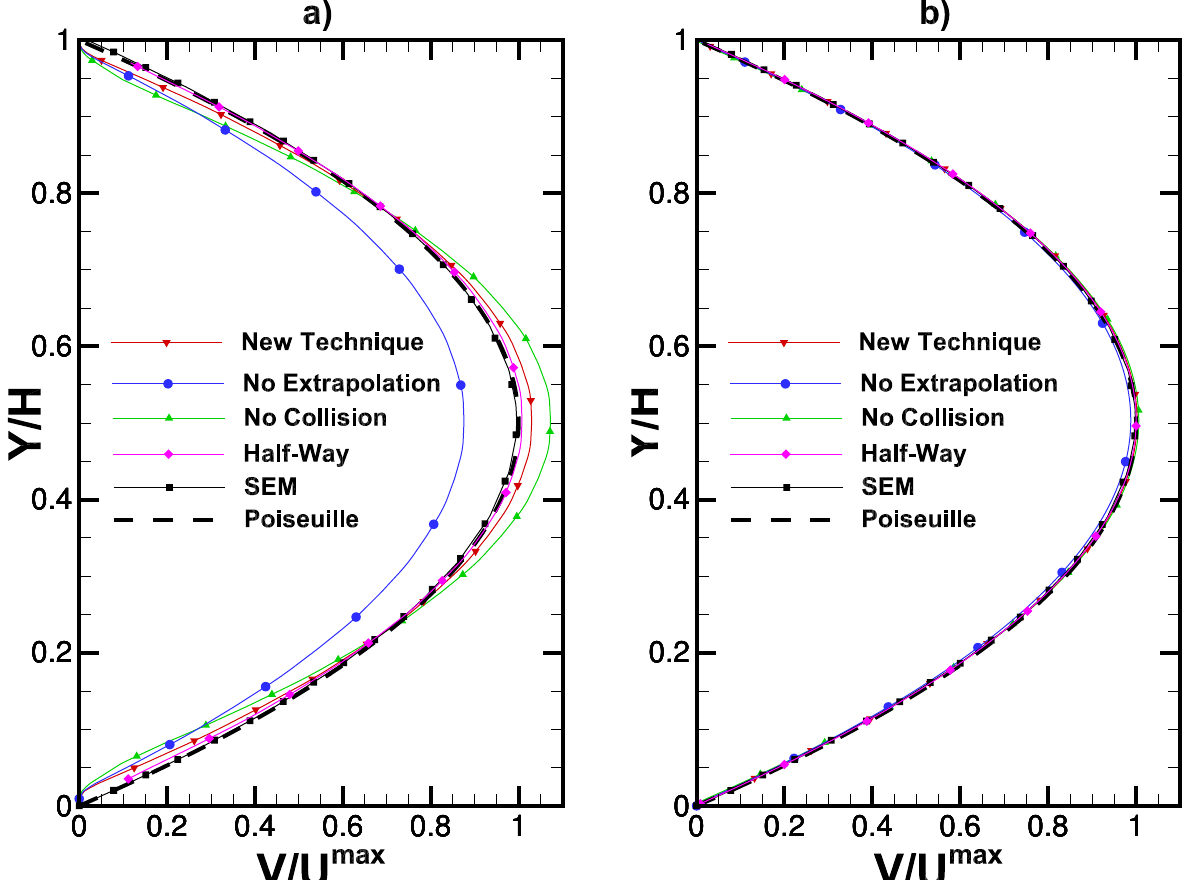}
} 
\caption{Velocity profiles at station 4: a) low grid resolution and Re = 0.01 and b) high grid resolution and Re = 1.}
  \label{fig:Inclined-Vel-profiles-Sland2}
  \end{center}
\end{figure}

\begin{figure}[h!]
  \begin{center}
    \includegraphics[width=0.4\textwidth,trim={2.0cm 2.5cm 2.5cm 2.0cm},clip]
    {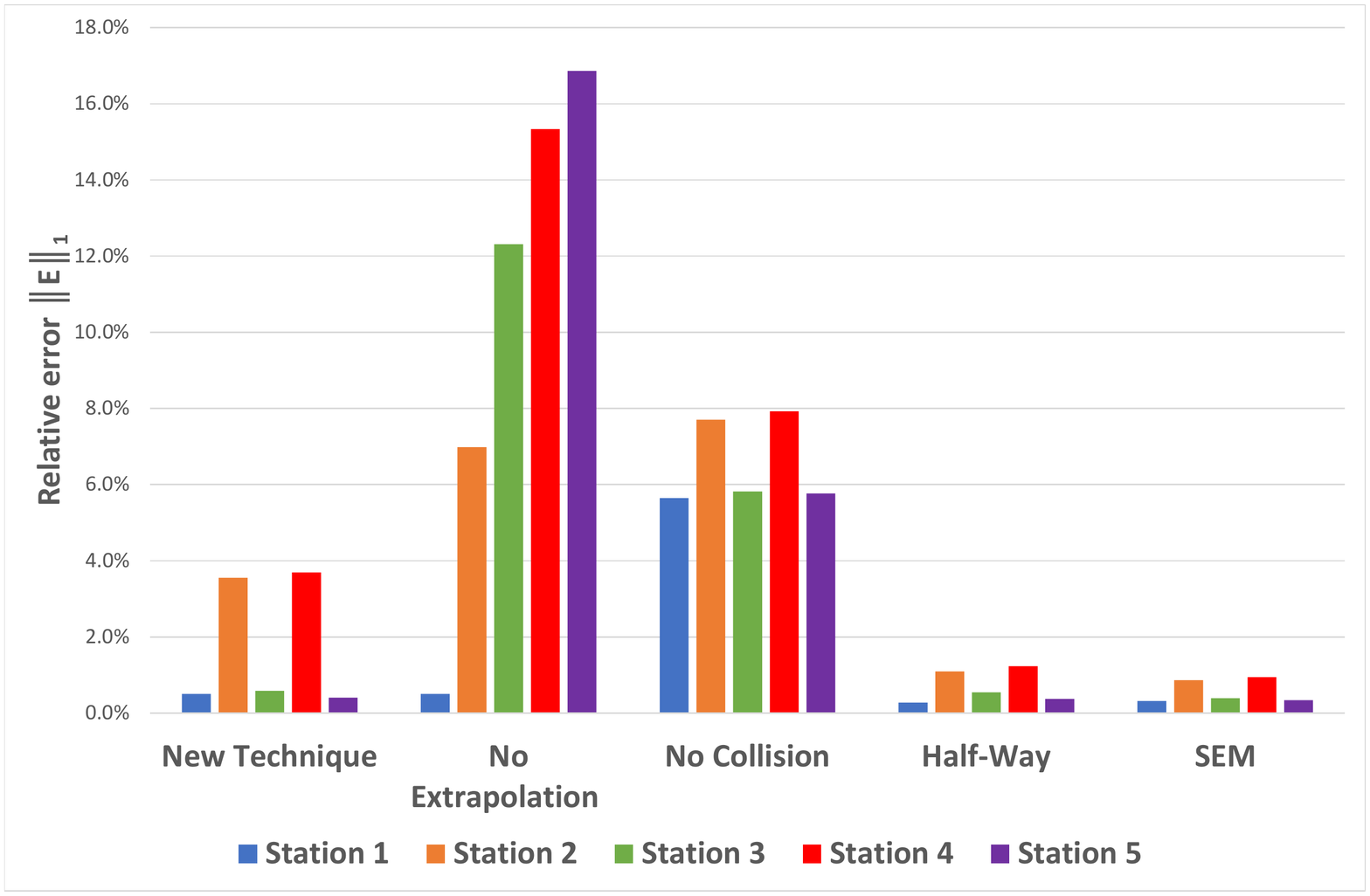} 
\caption{$\left\Vert E\right\Vert_1$ for low grid resolution and Re=0.01}
  \label{fig:Inclined-L1-LowReso-Re0-01}
  \end{center}
\end{figure}

\begin{figure}[h!]
  \begin{center}
    \includegraphics[width=0.4\textwidth,trim={2.0cm 2.5cm 2.5cm 2.0cm},clip]
    {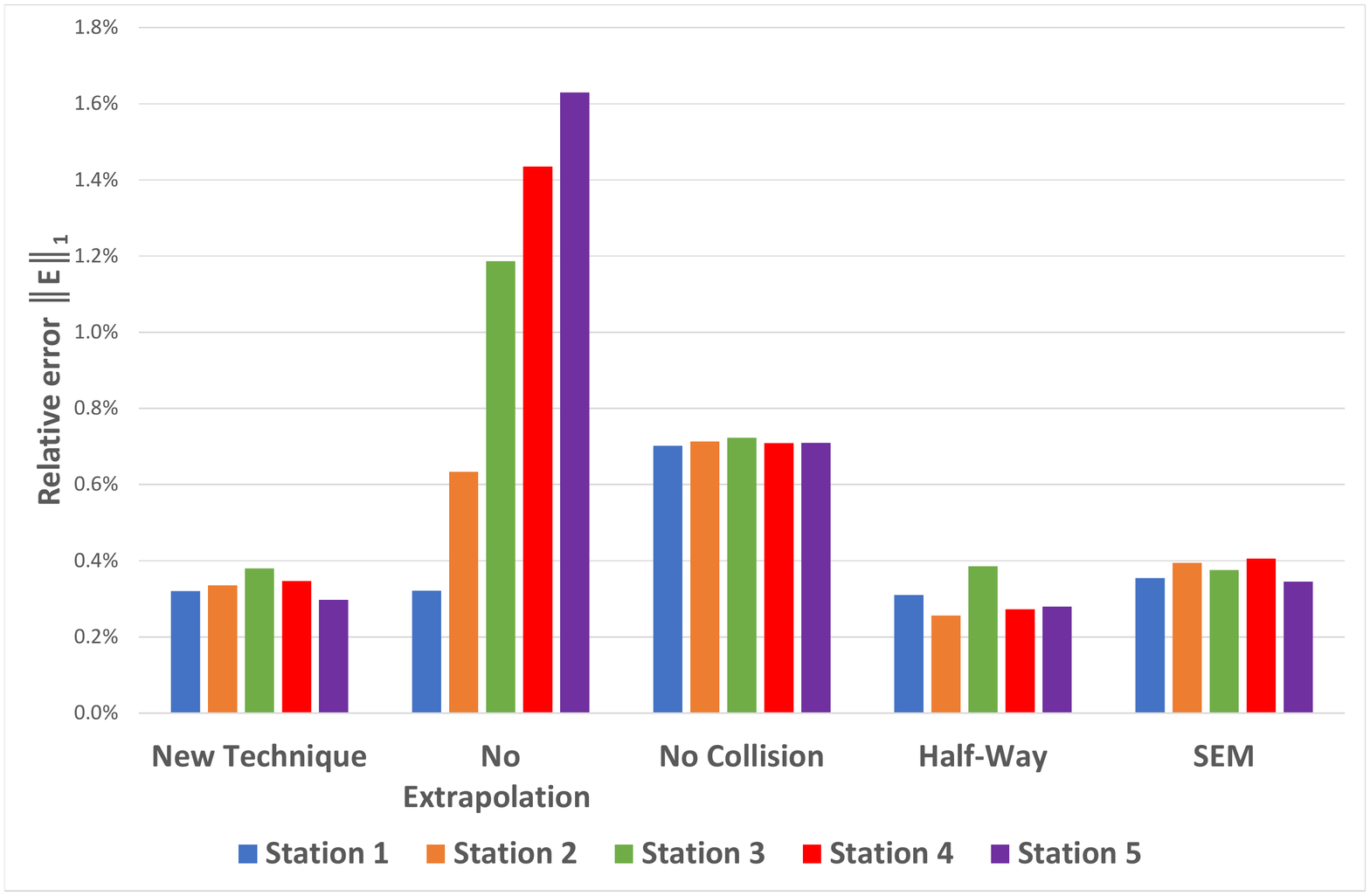} 
\caption{$\left\Vert E\right\Vert_1$ for high grid resolution and Re=0.01}
  \label{fig:Inclined-L1-HighReso-Re0-01}
  \end{center}
\end{figure}

\paragraph*{Velocity profiles}
In \autoref{fig:Inclined-Vel-profiles-Sland1}a, the velocity profiles at the station 2  for low grid resolution show that the ``Half-Way'' agrees with both SEM and Poiseuille profiles, but, the ``No Collision'' does not agree and has an overshoot at the centre line and undershoot at the wall. The ``No Extrapolation'' exhibits a loss of momentum. The ``New Technique'' agrees quite accurately, however, the heigh of the channel at the station 2 is smaller than H owing to inclined nature of the channel. The reduced height is due to the stair-like pattern used to represent inclined walls in this method. The effect of the geometry imposed constriction results in a slight overestimation of the velocity at the centre line in order to conserve momentum. Moreover, the deviation of the velocity at the wall comes from the no-slip condition imposed by the full-way bounce-back rule.  
\\
In \autoref{fig:Inclined-Vel-profiles-Middle}, the velocity profiles at station 3 is shown. This is the straight part at the middle of the geometry. The constriction effect due to the stair-like pattern does not influence this region i.e. the surface is aligned with the lattice. Thus, as seen in straight channel analysis, \autoref{fig:Inclined-Vel-profiles-Middle} shows a very good agreement between the ``New Technique'', ``SEM'', ``Half-Way'' and the analytical result. For the ``No Extrapolation'' and the ``No Collision'' shows inaccuracy.
\\
Similar results to station 2 are observed for station 4 with more loss of momentum for the ``No Extrapolation'' as shown in \autoref{fig:Inclined-Vel-profiles-Sland2}.
\paragraph*{Velocity profile errors}
Looking at the error norms based on the analytical solution i.e. a parabolic profile, the error is reduced by a factor 0.1 between the low resolution (\autoref{fig:Inclined-L1-LowReso-Re0-01}) and the high resolution (\autoref{fig:Inclined-L1-HighReso-Re0-01}) as expected i.e. linear convergence.

\begin{figure}[h!]
  \begin{center}
    \includegraphics[width=0.4\textwidth,trim={2.0cm 2.5cm 2.5cm 2.0cm},clip]
    {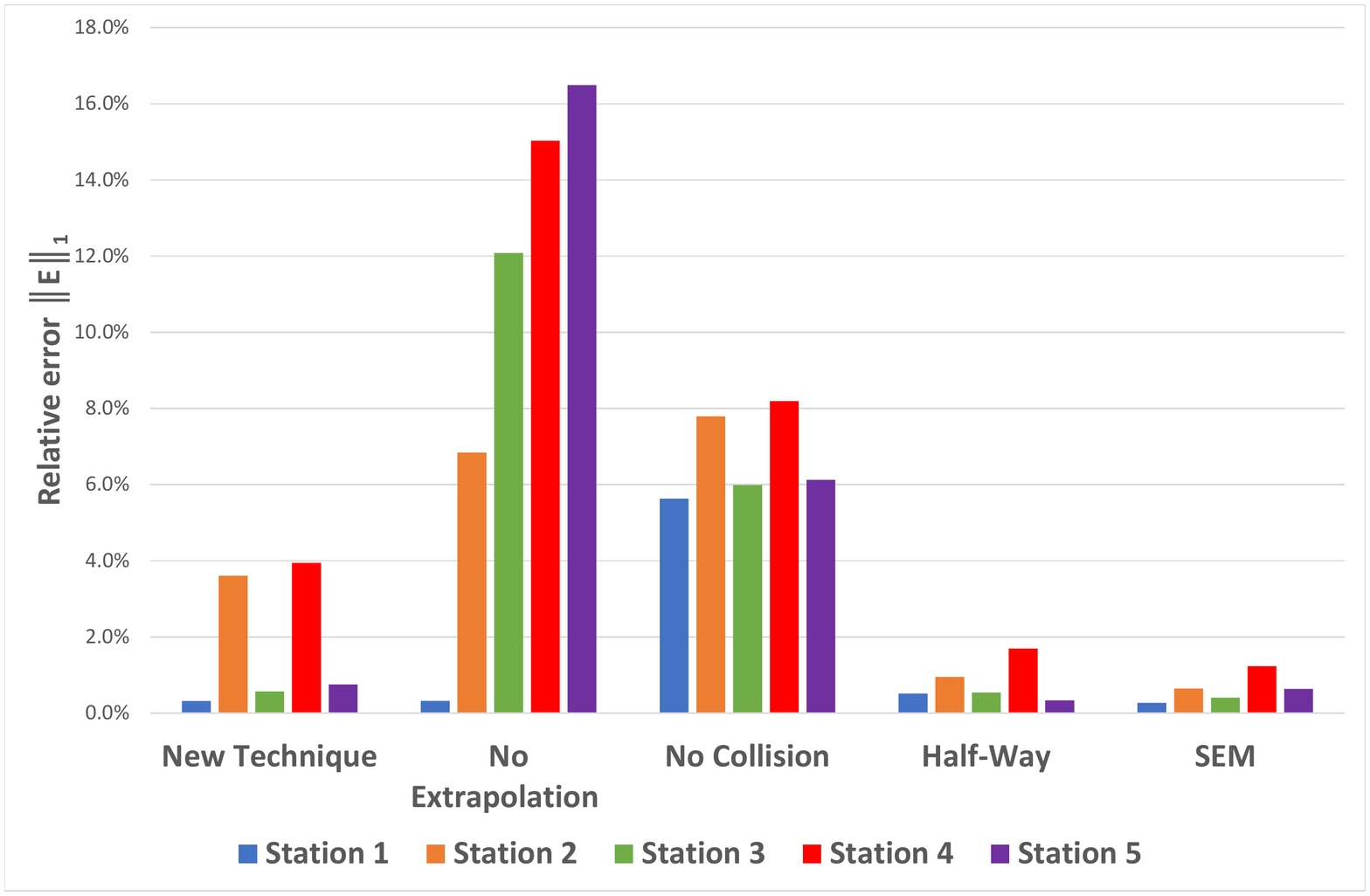} 
\caption{$\left\Vert E\right\Vert_1$ for low grid resolution and Re=1}
  \label{fig:Inclined-L1-LowReso-Re1}
  \end{center}
\end{figure}

\begin{figure}[h!]
  \begin{center}
    \includegraphics[width=0.4\textwidth,trim={2.0cm 2.5cm 2.5cm 2.0cm},clip]
    {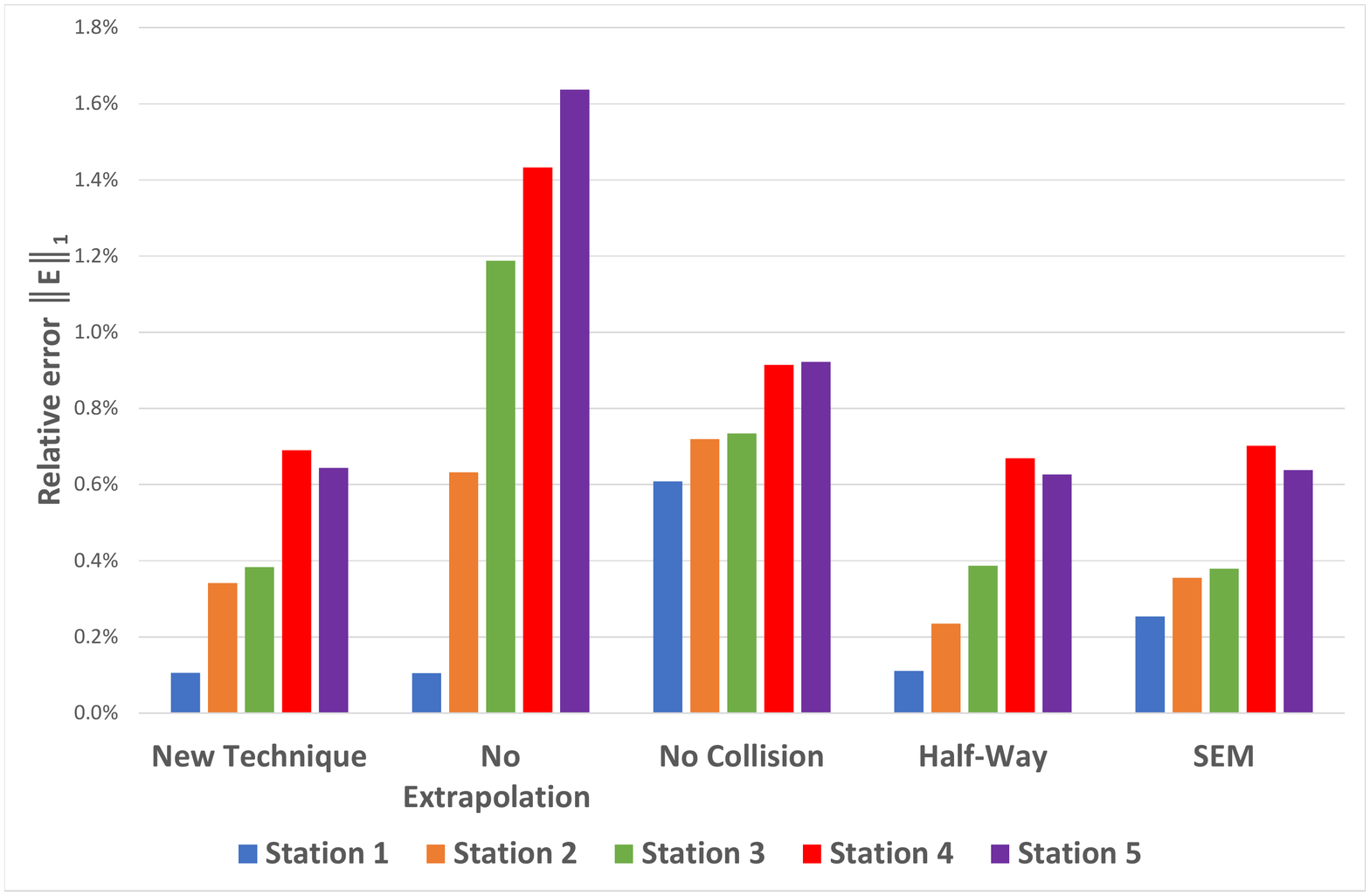} 
\caption{$\left\Vert E\right\Vert_1$ for high grid resolution and Re=1}
  \label{fig:Inclined-L1-HighReso-Re1}
  \end{center}
\end{figure}
The ``Half-Way'' and ``SEM'' results are essentially the same for Re= 0.01 or 1 and the low or high grid resolutions. The ``New Technique'' has some discrepancy for the inclined parts, however it is almost two times better than ``No Collision'' and four times better than ``No Extrapolation''.\\
At low grid resolution, the errors of each methods are similar for Re $\leqslant$ 1 (Figures \ref{fig:Inclined-L1-LowReso-Re0-01} and \ref{fig:Inclined-L1-LowReso-Re1}). However, the inertia starts to affect the solution at Re = 1 for the high grid resolution (\ref{fig:Inclined-L1-HighReso-Re0-01}, and\ref{fig:Inclined-L1-HighReso-Re1}). Note that we have also extracted the errors for Re=0.1 and the low or high grid resolutions with the similar results to Re=0.01.\\
The errors are accumulated for the ``No Extrapolation'' case due to the loss of momentum. We can notice the parabolic profile becomes less valid for Re=1 as shown in \autoref{fig:Inclined-L1-HighReso-Re1}.
\subsection{Effect on the pressure-drop}
\begin{figure*}[htbp!]
  \begin{center}%
    \includegraphics[width=0.8\textwidth,trim={0.6cm 1 0.6cm 0cm},clip]{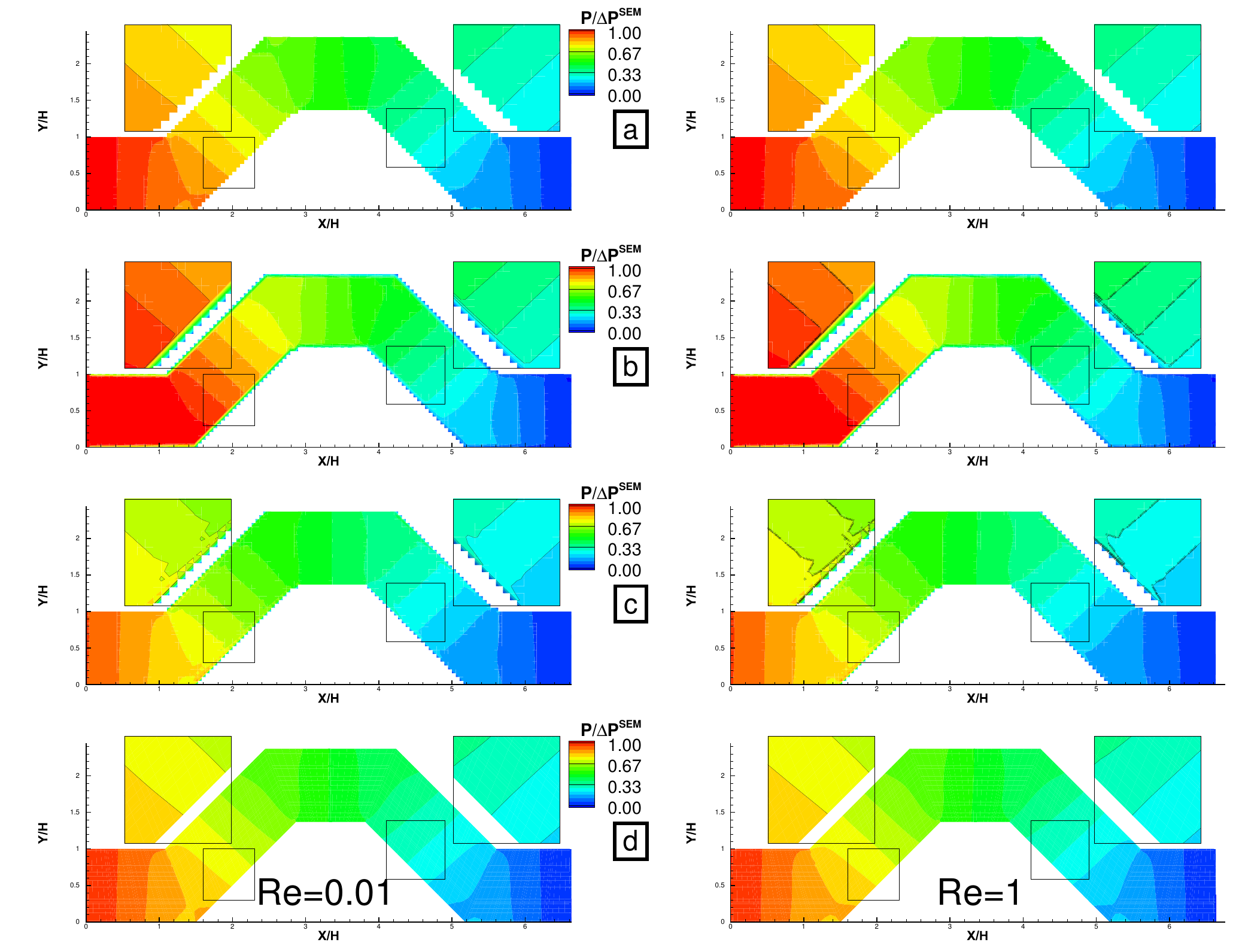} 
\caption{Comparison of pressure field to SEM results ; a) New Technique, b) No Collision, c) No Extrapolation, and d) SEM.}
  \label{fig:Inclined-Compare-Pressure}
  \end{center}
\end{figure*}

It is important to calculate the right pressure-drop, especially in low speed flows since at the incompressible limit, the velocity and density are directly coupled. Indeed, the  Navier-Stokes equations become elliptical. However, in two phases flow, the local pressure gradient is also important because this could change the shape and the displacement of an interface between two immiscible fluids.\\
This ``New Technique'' improves the prediction of the pressure-drop in case of low resolution. In \autoref{fig:Inclined-Compare-Pressure}, it can be clearly seen that the inaccurate value of density at the walls is predicted for the ``No Collision'' case. The ``No Extrapolation'' technique for the inclined parts yields very strange profiles (\autoref{fig:Inclined-Compare-Pressure}).\\
\\
In \autoref{Tab:Poiseuille-Inclined}, the pressure-drop is compared to the results from SEM. It can be clearly seen that the pressure-drop is not correctly calculated in case of ``No Collision'' while the results of the ``New Technique'' and SEM are in good agreement. It can be noticed that the ``Half-Way'' technique cannot converge under 1\% since the shear stress is taken in account at half lattice from the wall which leads to errors when the flow is not parallel to the lattices.
\begin{table}[htbp!]
\begin{ruledtabular}
\begin{tabular}{cccccc}
\multicolumn{6}{c}{\textbf{Inclined channels}}                                                                                        \\ \hline
\multicolumn{1}{l}{} & \textbf{Re} & \thead{\textbf{New} \\ \textbf{Technique}} & \thead{\textbf{No} \\ \textbf{Extrapolation}} & \thead{\textbf{No} \\ \textbf{Collision}} & \thead{\textbf{Half-} \\ \textbf{Way}} \\ \hline
\multirow{4}{*}{\thead{\textbf{Low}\\ \textbf{Resolution}}} 
& 10    & 12.20\%    & 2.74\%      & 29.59\%      & 3.45\%   \\ 
& 1     & 6.17\%     & 4.90\%      & 21.04\%      & 7.85\%   \\ 
& 0.1   & 5.66\%     & 5.08\%      & 20.32\%      & 8.22\%   \\ 
& 0.01  & 5.61\%     & 5.10\%      & 20.25\%      & 8.25\%   \\ \hline
\multirow{4}{*}{\thead{\textbf{High} \\ \textbf{Resolution}}} 
& 10    & 0.36\%     & 1.67\%      & 0.91\%      & 1.01\%  \\ 
& 1     & 0.31\%     & 1.57\%      & 0.70\%      & 1.01\%  \\ 
& 0.1   & 0.30\%     & 1.56\%      & 0.68\%      & 1.01\%  \\ 
& 0.01  & 0.28\%     & 1.53\%      & 0.68\%      & 1.18\%  \\ 
\end{tabular}
\caption{Pressure-drop errors for a Poiseuille-like flow in the inclined channels.}
\label{Tab:Poiseuille-Inclined}
\end{ruledtabular}
\end{table}
\\
Moreover, we notice the error without collisions on walls increases compared to a straight channel. This suggests that for a single-phase flow with a very complex geometry such as porous media, the effect of the collision could be not negligible in a single-phase flow.
\section{Conclusion}
Using the inverse distance weighting extrapolation on concave corners and the collisions on all walls nodes enable the new technique to conserve the momentum and improve the accuracy compared to the full-way bounce-back. However, this technique needs to be restricted to low Reynolds number since the pressure gradient is not directly taken into account i.e. the convection time scale is neglected compared to the diffuse time scale. The grid resolution can be also reduced, thus a bigger domain becomes affordable. We have not noticed an increase of the calculation cost due to the extrapolation but only due to the collision, even for a simulation on a Berea sandstone sample \cite{Boek2010} where, the domain was 1323 [lu] by 1059 [lu] with a huge number of walls.
\begin{acknowledgments}
We thank F. Donbosco for his useful comments and his global improvement of the quality of the paper. Results were obtained using the UK Consortium on Mesoscale Engineering Sciences (www.ukcomes.org) and the EPSRC funded ARCHIE-WeSt High-Performance Computer. The EPSRC grant numbers are No. EP/L00030X/1 and No. EP/K000586/1. The simulations ran on ARCHER (www.archer.ac.uk/) and ARCHIE-WeSt (www.archie-west.ac.uk).
\end{acknowledgments}

\newpage
\section*{References}
\bibliography{mybibfile}

\end{document}